\documentclass[aps,showpacs,twocolumn]{revtex4}
\usepackage{graphicx}
\def\bra{\langle}
\def\ket{\rangle}
\def\<{\langle}
\def\>{\rangle}

\begin{document}

\title{Distinctness of ensembles having the same density matrix\\
and the nature of liquid NMR quantum computing}

\author{Gui-Lu Long$^{1,2}$, Yi-Fan Zhou$^1$,
Jia-Qi Jin$^{1}$, and Yang Sun$^{3,1}$}

\affiliation{$^1$ Key Laboratory For Quantum Information and
Measurements and Department of Physics, Tsinghua University,
Beijing $100084$, P.R. China\\
$^2$ Center for Atomic and Molecular NanoSciences, Tsinghua
University, Beijing 100084, P.R. China\\
$^3$ Department of Physics, University of Notre Dame, Notre Dame,
Indiana 46556, USA }

\date{\today}

\begin{abstract}
We show that differently constructed ensembles having the same
density matrix may be physically distinguished by observing
fluctuations of some observables. An explicit expression for
fluctuations of an observable in an ensemble is given. This result
challenges Peres's fundamental postulate and seems to be contrary
to the widely-spread belief that ensembles with the same density
matrix are physically identical. This leads us to suggest that the
current liquid NMR quantum computing is truly quantum-mechanical
in nature.
\end{abstract}

\pacs{03.67.Dd, 03.67.Hk, 03.67.-a}
\maketitle

States of an ensemble can be described by a density matrix. Unlike
a pure state, a mixed state may be prepared in many different
ways. Differently-prepared ensembles may have the same density
matrix. It is commonly accepted that ensembles having the same
density matrix (ESD) can not be distinguished by any conceivable
measurements \cite{r1,r2,r3}, and hence should be considered as
physically identical. Peres \cite{r13} expressed this as a
postulate: ``The $\rho$ matrix completely specifies all the
properties of a quantum ensemble." However, some other authors
expressed their caution. Fano used preparation history to specify
an ensemble \cite{r4}. d'Espagnat emphatically pointed out that
different ensembles having the same density matrix may be
physically distinguished, and in Ref. \cite{r5}, he gave a
concrete example for a qubit ensemble. Apparently, d'Espagnat's
point of view has not been widely noticed as many in the community
tend to believe that ESD can not be physically distinguished.
Clarifying this issue has become extremely important since in
recent years the concept of mixed state has been extensively used
in quantum computation and quantum communication. The prevalent
understanding on ESD has already led to serious questions
\cite{r6} about the nature of quantum computation: the liquid
nuclear magnetic resonance quantum computation (NMR QC) is not
truly quantum-mechanical because of lack of entanglement.

The question by Braunstein {\it et al.} \cite{r6} was raised upon
rewriting the effective pure state density matrix in terms of a
convex decomposition of density matrices of product states
\begin{eqnarray}
\rho=(1-\epsilon)M_d+\epsilon\rho_{\Delta}.
\label{e1}
\end{eqnarray}
In Eq. (\ref{e1}), $\epsilon$ is called the polarization and $M_d$
is a $d\times d$ unit matrix divided by $d$. They then concluded
that if $\epsilon$ in the equation is sufficiently small, the
effective pure state in NMR QC can be replaced by an ensemble with
separable states, and therefore, there is no entanglement. They
thus suggested \cite{r6} that the current NMR QC should only be
considered as a classical simulation but not as a true quantum
computation. Though Laflamme \cite{r7} did not agree with the
conclusion of Braunstein {\it et al.} in \cite{r6}, he did agree
that in each step of the present NMR QC experiment the state has
no quantum entanglement. However, there is one puzzle: if one
believes that without entanglement, NMR QC is still more powerful
than a classical computing, then what is the source of the NMR QC
power, as DiVincenzo questioned \cite{r8}?

In Ref. \cite{r9}, some of us attempted to understand this
subject. The main points were: 1) The density matrix describes the
state of ``an average particle" in an ensemble. It does not
describe the state of any individual particle in an ensemble; 2)
Entanglement is a property of the wave function of an individual
particle, such as a molecule in a liquid NMR sample. Separability
of the density matrix can not be used to measure entanglement of
an ensemble; 3) The state evolution in a bulk-ensemble NMR
computation is a quantum process; 4) The coefficient $\epsilon$ in
Eq. (\ref{e1}) is a measure of simultaneity of the molecules in an
ensemble, which reflects the intensity of the NMR signal and has
no significance in quantifying the entanglement in a bulk ensemble
NMR system. Based on these arguments, it was suggested \cite{r9}
that the current NMR QC is truly quantum-mechanical.

In this Letter, we extend the study by pointing out explicitly
that the conclusion that ESD are physically identical, which is a
form of Peres's postulate and the pillar stone of Ref. \cite{r6}
that has led to doubts on the quantum nature of the current NMR
computation, is not general. We begin with an examination of the
mixed state concept. Then we focus our discussion on the so-called
proper mixed state that describes the state of an ensemble. We
show how ESD can be distinguished by considering fluctuations of
some observables. We finally come to the question about the nature
of quantum computing, and conclude that the current NMR QC is
genuinely quantum-mechanical.

A mixed state is the description of state of a quantum system ``if
we do not even know what state is actually present -- for example,
when several states $\phi_1$, $\phi_2$,... with respective
probabilities $w_1$, $w_2$, $\cdots$ ($w_1\geq0$, $w_2\geq 0$,
$\cdots$, $w_1+w_2+\cdots=1$) constitute the description"
\cite{r10,r11}. {\it Operationally, a mixed state is usually
associated with certain averaging processes.} There are three ways
to obtain a mixed state. {\bf The first way}, which was also used
by von Neumann in his original papers \cite{r10,r11}, is to
average over an ensemble of particles. For simplicity we call
these particles molecules. Suppose an ensemble contains $N$
molecules, with $Nw_1$ molecules in state $|\phi_1\ket$, $Nw_2$
molecules in state $|\phi_2\ket$, and so on. Then the state of an
averaged particle in this ensemble is a mixed state, described by
the density matrix
\begin{eqnarray*}
\rho=w_1|\phi_1\ket\bra\phi_1|+w_2|\phi_2\ket\bra\phi_2|+\cdots
.\label{e2}
\end{eqnarray*}
d'Espagnat referred this kind of mixed state to as a {\it proper
mixed state} \cite{r5}.

{\bf The second way} is to make the averaged state of a molecule
entangled with other degrees of freedom. By averaging over the
other degrees of freedom using the trace operation, the reduced
density matrix may become the one representing a mixed state. For
example, for a quantum system containing two particles $A$ and $B$
in state $|\phi_{AB}\ket$, or equivalently
$\rho_{AB}=|\phi_{AB}\ket\bra \phi_{AB}|$, the reduced density
matrix for particle $A$ is
\begin{eqnarray}
\rho_A={\rm Tr}_B(\rho_{AB}).\label{e3}
\end{eqnarray}
The mixed state arising from this way is called by d'Espagnat as
an {\it improper mixed state}. d'Espagnat stressed that an
improper mixed state can not be associated with a proper mixed
state with the same density matrix \cite{r5}.

{\bf The third way} is to obtain the averaged state of a particle
over a random variable, or over a period of time. For instance,
when a spin-$1\over 2$ particle is subject to a random kick
interaction of the form $ K=\exp[i\sigma_z\theta/\hbar]$, where
$\theta$ is a random variable with distribution $P(\theta)$, the
averaged state of this particle over the random variable,
$\bra\rho\ket=\int_{-\infty}^{+\infty}P(\theta) \rho'd\theta$,
normally gives a mixed state, albeit at any instant the particle
is in a pure state.

In the following we consider mainly the proper mixed state. The
expectation value of observable $A$  for an averaged molecule in
an ensemble is
\begin{eqnarray} \bra \hat{A}\ket ={\rm Tr}( \rho A),
\label{e4}
\end{eqnarray}
and expectation value of $A$ for the whole ensemble of $N$
molecules is
\begin{eqnarray}
\bra \hat{A}\ket_E=N\bra \hat{A}\ket=N{\rm
Tr}(\rho A).
\label{e5}
\end{eqnarray}
From the expectation values in (\ref{e4}) and (\ref{e5}), ESD can
of course not be distinguished. However, this does not exclude the
possibility of distinguishing them with other physical means.
d'Espagnat has given an explicit example. Ensemble $S_I$ is
prepared by putting $N/2$ qubits in state $|z\ket=|0\ket$ and
another $N/2$ qubits in state $|-z\ket=|1\ket$. In ensemble
$S_{II}$, $N/2$ particles are put in state
$|+x\ket=(|0\ket+|1\ket)/\sqrt{2}$, and another $N/2$ particles in
state $|-x\ket=(|0\ket-|1\ket)/\sqrt{2}$. The density matrices for
the two ensembles can be written as
\begin{eqnarray}
S_I&=&{1\over 2}(|0\ket\bra 0|+|1\ket\bra 1|),\label{sz}\\
S_{II}&=&{1\over 2}(|+x\ket\bra +x|+|-x\ket\bra -x|)
\label{sx}.
\end{eqnarray}
These ensembles are equivalent to those in the example used by
Peres to show the indistinguishability of ESD \cite{r14}. In the
matrix form, they both lead to the same matrix: $M_2$. However,
d'Espagnat used the fluctuation of the 3rd component of the spin
operator
\begin{eqnarray}
\Sigma_z=\sum_{i=1}^N\sigma_z(i),\label{sigmaz}
\end{eqnarray}
where the summation is over all molecules in the ensemble, to
distinguish the two ensembles. For ensemble $S_I$, the fluctuation
$\left(\Delta\Sigma_{I,z}\right)_E=0$, and for ensemble $S_{II}$,
$\left(\Delta\Sigma_{II,z}\right)_E=\sqrt{N}$.

Preskill \cite{r15} also noticed that the two ensembles can be
physically distinguished by using a different method. In his
example, Alice and Bob share $N$ pairs of particles in state
$|\psi_{AB}\ket={1\over
\sqrt{2}}(|0\ket_A|0\ket_B+|1\ket_A|1\ket_B)$. Bob measures his
particles in either $\sigma_z$ or $\sigma_x$, and Alice's
particles is then prepared in either $S_I$ or $S_{II}$. Bob tells
Alice of the result of his measurement for each particle, but does
not tell her the measured apparatus. However, Alice can
distinguish the two ensembles with the information at hand: Alice
measures each of the particles in her ensemble using the
$\sigma_z$-basis and compares her result with Bob's. If her result
has a perfect agreement with Bob's, then she knows that the
ensemble is $S_I$. Otherwise it is the ensemble $S_{II}$. Hence
Alice ``does have a way to distinguish Bob's two preparation
methods" \cite{r15}.

It has become clear from the above discussion that different
preparations of ensembles of qubits can be distinguished
physically. We now generalize d'Espagnat's idea and show that
there exists a general formalism to distinguish ESD. By measuring
fluctuations, ensemble preparation can be distinguished up to
composition: the number of particles in each of the pure state in
an ensemble. We consider different preparations of an ensemble
leading to the same composition as the same. For instance,
preparing $N_1$ molecules first in steps of $N_1-3$ and then
adding another 3, or in two equal steps with $N_1$/2 each, will be
considered as the same. For an ensemble with $N$ molecules with
$N_i$ molecules in a state $|\psi_i\ket$, the density matrix is
\begin{eqnarray}
\rho={N_1\over N}|\psi_1\ket\bra \psi_1|+{N_2\over
N}|\psi_2\ket\bra \psi_2|+\ldots+{N_m\over N}|\psi_m\ket\bra
\psi_m|,\label{statem}
\end{eqnarray}
where $m$ is the number of pure states within the ensemble.
Considering an observable $\Omega$ for an ensemble
\begin{eqnarray}
\Omega=\sum_{i=1}^N\Omega(i),
\end{eqnarray}
where $\Omega(i)$ is the observable for molecule $i$. The
expectation value of $\Omega$ for this ensemble can be calculated
according to Eq. (\ref{e5}). The fluctuation of $\Omega$ is
\begin{eqnarray}
\Delta\Omega_E&=&\sqrt{\sum_{i=1}^mN_i\left(\bra\psi_i|\Omega^2|\psi_i\ket
-\left(\bra\psi_i|\Omega|\psi_i\ket\right)^2\right)}\nonumber\\
&=&\sqrt{N{\rm
Tr}(\rho\Omega^2)-\sum_iN_i(\bra\psi_i|\Omega|\psi_i\ket)^2}.\label{fluc}
\end{eqnarray}
The term $\sum_iN_i(\bra\psi_i|\Omega|\psi_i\ket)^2$ is sensitive
to the ensemble composition. Here, $\Omega(i)$ is not restricted
to one-body operators; for molecules that contain composite
constituents it can usually be many-body operators. For instance,
the two-body operator $\Sigma_{zz}$ in an ensemble of molecules
with 7 qubits can be written as
\begin{eqnarray}
\Sigma_{zz}=\sum_{i=1}^N\left(\sum_{a<b=1}^7
\sigma_{a,z}(i)\sigma_{b,z}(i)\right).
\end{eqnarray}
By choosing different observables $\{\Omega, \Lambda,\cdots\}$,
one can then distinguish different ESD.

In principle, the state of an ensemble of $N$ molecules with
composition (\ref{statem}) can be written as
\begin{eqnarray}
|\psi\ket=\overbrace{|\psi_1\ket_1\cdots|\psi_1\ket_{N_1}}\;
\overbrace{|\psi_2\ket_{N_1+1}\cdots|\psi_2\ket_{N_1+N_2}}\cdots,
\label{staten}
\end{eqnarray}
though we usually do not have the detailed information about which
molecule is in what state. We stress here that molecules in an
ensemble are classically distinguishable, and that this is
different from the state in a Bose-Einstein condensate where the
particles are indistnguishable. The ensemble results (\ref{e5})
and (\ref{fluc}) can also be calculated directly by using Eq.
(\ref{staten}), which leads to the same conclusion.

Now we show that the ensemble of the effective Bell-state and its
product state decomposition presented in Ref. \cite{r6} can be
physically distinguished. The effective Bell-state density matrix
is
\begin{eqnarray}
\rho_1=(1-\epsilon)M_4+\epsilon\rho_{Bell},
\label{rhobell}
\end{eqnarray}
where $\rho_{Bell}=(|00\ket+|11\ket)(\bra 00|+\bra 11|)/2$. The
effective Bell-state describes an ensemble with $\epsilon N$
molecules in Bell-state $(|00\ket+|11\ket)/\sqrt{2}$ and
$(1-\epsilon)N/4$ molecules in each of the calculating-basis
states $|00\ket$, $|01\ket$, $|10\ket$ and $|11\ket$. According to
Ref. \cite{r6}, it can be decomposed into
\begin{eqnarray}
\rho_2=\sum_{i,j}{1\over 4}\left({1\over
9}+C_{ij}\right)P_i\otimes P_j, \label{rhobell2}
\end{eqnarray}
where $P_i=(1\mp\sigma_i)/2$ represents a pure state polarized or
anti-polarized along the the three axes, $x$, $y$, and $z$,
respectively. The two different compositions of the effective
Bell-state can be distinguished by observing fluctuations of the
observable $ \Sigma_{zz}=\sum_{i}\sigma_{1z}(i)\sigma_{2z}(i)$.
For the ensemble (\ref{rhobell}), the fluctuation is
$(\Delta\Sigma_{zz})_{1,E}=\epsilon \sqrt{N}$, whereas for the
product state expansion (\ref{rhobell2}), the fluctuation is
$(\Delta\Sigma_{zz})_{2,E}={2\sqrt{N} \over 3}$. Hence the two
ensembles are distinguished though they both have the same density
matrix.

The decomposition of the effective Bell-state into a product
states can not be used to infer the conclusion that the effective
Bell-state is not entangled because entanglement is the property
of an individual molecule, and not of an ensemble. We would like
to point out another inconsistent problem in Ref. \cite{r6} in the
description of the product state decomposition for NMR QC. In the
effective pure state NMR QC, the QC computation is performed on
$\epsilon N$ molecules, with each molecule working as a single
quantum computer. They go through one state to another in a
computing process. However, in the product state decomposition,
once an entangling operation such as the controlled-NOT gate is
performed, the number of particles within each composition will
change.

If each molecule in an ensemble is in a mixed state itself, even
at a given instant, there are fluctuations of an observable
$\Omega$ \cite{r16,r17}
\begin{eqnarray}
\Delta\Omega_E=\sqrt{N{\rm Tr}(\rho\Omega^2)-N({\rm
Tr}(\rho\Omega)^2}.
\end{eqnarray}
In such a case, all ensembles are equivalent and no physical means
can distinguish them. In fact, there is no difference at all among
them because the compositions of the ensembles are the same, and
they all are made up by molecules in the same mixed state. This
fact has been used by some authors to criticize d'Espagnat
\cite{r18}. To this objection, we stress that it is impossible to
prepare such an ensemble using classical probabilities. For
instance, in the example given by Peres \cite{r15}, the state of a
photon is prepared according to the result of a coin-tossing,
either along $z$ or along $-z$ axis. It is true that each photon
has 50\% probability to be in state $|0\ket$ or $|1\ket$. But each
photon is prepared in a definite state, not in a mixed state.
Similarly, in the BB84 quantum key distribution protocol, the
state of a qubit has 25\% probability in each of the four states
$|z\ket$, $|-z\ket$, $|x\ket$, and $|-x\ket$. But it is certainly
in one of the four possible states. This ensemble has the density
matrix $M_2$. The same density matrix can be realized by preparing
each photon with 50\% probability in $|z\ket$ and $|-z\ket$,
respectively. Though these two different preparations result in
the same density matrix, their security aspect is totally
different: the BB84 protocol is unconditionally secure whereas the
one based on the latter is completely insecure.

One possible way to realize such a scenario is that each molecule
is actually in an improper state. Suppose that we have an ensemble
of two-qubits molecules, and the two qubits are in an entangled
state, say
\begin{eqnarray}
|\psi_{AB}\ket_1={1\over
\sqrt{2}}\left(|00\ket_{AB}+|11\ket_{AB}\right).
\end{eqnarray}
If we look at the first qubit in each molecule in the ensemble,
the resulting state is an ensemble of molecules, and each molecule
is in an improper mixed state. Then in this case, by observing the
observables of the qubit $A$ alone one can not distinguish this
ensemble from another ensemble if the two-qubits molecules are all
in the state
\begin{eqnarray}
|\psi_{AB}\ket_2={1\over
\sqrt{2}}\left(|01\ket_{AB}+|10\ket_{AB}\right).
\end{eqnarray}
However, we know by now that by looking at the fluctuations of a
two-body operator, for example,
$\sum_{i=1}^N\sigma_{A,z}(i)\sigma_{B,z}(i)$, one can distinguish
between these two ensembles.

Which picture is closer to the liquid NMR QC, an ensemble of
molecules with each molecule in a definite pure state at a given
instant, or an ensemble with every molecule in the same mixed
state? We favor the first picture. In the first picture, at a
given instant, every molecule in the NMR sample is in a pure state
and each remains in the pure state for some time before changing
to another pure state. In a long time scale, each molecule in an
ensemble is in a mixed state.  The spin-relaxation time $T_1$ in
NMR can be viewed as the time period in which a molecule remains
in a certain quantum state. This is supported by the Gorter
formula \cite{r20}
\begin{eqnarray}
{1\over T_1}={1\over 2} {\sum_{n,m}W_{n,m}(E_m-E_n)^2 \over \sum_n
E_n^2},
\end{eqnarray}
where $W_{m,n}$ is the transition probability rate from level $m$
to level $n$. Hence $T_1$ is the energy weighted time in which a
molecule remains in a given quantum state. A liquid NMR QC is
completed within the decoherence time, which is much shorter than
$T_1$. During a NMR QC, a molecule is well approximated as being
in a definite quantum state. This picture has been used in the
effective NMR QC literatures. Gershenfeld and Chuang \cite{r19}
have explicitly explained this in an effective pure state in which
there are some numbers of molecules in a definite quantum state.
In addition, nuclear spins are well isolated from the environment;
its entanglement with the environment is weak. For those nuclear
spins that are not considered in quantum computing, decoupling
pulses are usually used to disentangle them from the working
nuclear spins so that they can not form an entangled state that
will make the working qubits in an improper mixed state.

In summary, we have shown that for ensembles of molecules with the
same density matrix, fluctuations can be used to distinguish
different ensemble compositions. This result is a direct
generalization of d'Espagnat, and it challenges one of Peres'
fundamental postulates and corrects the common belief that ESD can
not be physically distinguished. The direct consequence of this
result is the invalidation of the conclusion in Ref. \cite{r6}
where the authors introduced the decomposition of the effective
pure state into product states to claim the absence of
entanglement in the current NMR QC experiment. We emphasize that
entanglement is the property of individual molecules within an
ensemble, not of the whole ensemble. We suggest that during an NMR
QC experiment, quantum entanglement does exist in individual
molecules, and this is exactly the source of enhancement of
computing power in NMR QC. As the only venue capable of
demonstrating quantum algorithms at present, NMR QC is a useful
test-bed for quantum computing. Recently, the para-hydrogen
technique has been used to enhance the polarization to near unity
\cite{r21}. This makes the liquid NMR continuing to play the
unique role in quantum information studies.

This work is supported by the National Fundamental Research
Program, Grant No. 001CB309308, China National Natural Science
Foundation, Grant No. 60073009, 10325521, the SRFDP program of
Education Ministry of China.

\end{document}